\documentclass[aps,prb,twocolumn,superscriptaddress]{revtex4-1}
\usepackage{graphicx}
\usepackage{color}
\usepackage{subfigure}
\usepackage{setspace}

\begin{document}


\title{Burning and graphitization of optically levitated nanodiamonds in vacuum}


\author{A T M A. Rahman}
\email{a.rahman@ucl.ac.uk}
\affiliation{Department of Physics and Astronomy, University College London, Gower Street,  WC1E 6BT, UK}
\affiliation{Department of Physics, University of Warwick, Gibbet Hill Road, CV4 7AL, UK}
\author{A. Frangeskou}
\affiliation{Department of Physics, University of Warwick, Gibbet Hill Road, CV4 7AL, UK}
\author{M. S. Kim}
\affiliation{QOLS, Blackett Laboratory, Imperial College London, SW7 2BW, UK}
\author{S. Bose}
\affiliation{Department of Physics and Astronomy, University College London, Gower Street,  WC1E 6BT, UK}
\author{G. W. Morley}
\affiliation{Department of Physics, University of Warwick, Gibbet Hill Road, CV4 7AL, UK}
\author{P. F. Barker}
\email{p.barker@ucl.ac.uk}
\affiliation{Department of Physics and Astronomy, University College London, Gower Street,  WC1E 6BT, UK}



\date{\today}

\begin{abstract}
A nitrogen-vacancy (NV$^-$) center in a nanodiamond, levitated in high vacuum, has recently been proposed as a probe for demonstrating mesoscopic center-of-mass superpositions \cite{Scala2013, Zhang2013} and for testing quantum gravity \cite{Albrecht2014}. Here, we study the behavior of optically levitated nanodiamonds containing NV$^-$ centers at sub-atmospheric pressures and show that while they burn in air, this can be prevented by replacing the air with nitrogen. However, in nitrogen the nanodiamonds graphitize below $\approx 10$ mB. Exploiting the Brownian motion of a levitated nanodiamond, we extract its internal temperature ($T_i$) and find that it would be detrimental to the NV$^-$ center's spin coherence time \cite{Toyli2012}. These values of $T_i$ make it clear that the diamond is not melting, contradicting a recent suggestion \cite{Neukirch2015}. Additionally, using the measured damping rate of a levitated nanoparticle at a given pressure, we propose a new way of determining its size.
\end{abstract}


\maketitle

Even though diamond is thermodynamically metastable in ambient conditions, it has extremely high thermal conductivity, Young's modulus, electrical resistivity, chemical stability, and optical transparency \cite{Mochalin2012,Khmelnitsky2014,Davies1972,Shenderova2002}. Nanodiamonds inherit most of these spectacular properties from their bulk counterparts and the inclusion of color centers such the NV$^-$center has increased their realm of applications \cite{Doherty2013,Mochalin2012}. Proposed and demonstrated applications of diamond, nanodiamonds and nanodiamonds with NV$^-$ centers include tribology \cite{Ivanov2010,Mochalin2012}, nanocomposites \cite{Behler2009}, UV detection in space applications \cite{Pace2003}, magnetometry \cite{Taylor2008}, biological imaging \cite{Balasubramanian2008}, quantum information processing \cite{Dutt2007,Fuchs2011} and thermometry \cite{Toyli2012}. More recently nanodiamonds with NV$^-$ centers have been suggested for testing quantum gravity \cite{Albrecht2014} and for demonstrating center of mass (CM) superpositions of mesoscopic objects \cite{Scala2013, Zhang2013}. These superpositions and interferometry also point towards a broader future application of levitated diamonds in sensing and gravitometry. In the scheme for testing quantum gravity, an NV$^-$ center in a nanodiamond is exploited in a Ramsey-Borde interferometer \cite{Albrecht2014} and, in the nonrelativistic limit, the first order correction to the energy dispersion scales with the size of a nanodiamond. In the case of creating CM superpositions, the NV$^-$ center's spin is utilized and the spatial separation of the superposed CM states depends on the size of a nanodiamond \cite{Scala2013, Zhang2013}. To prevent the adverse effects of motional decoherence, these proposals \cite{Albrecht2014,Scala2013, Zhang2013} have been conceptualized in high vacuum ($10^{-6}$ mB). It is, however, well known that at atmospheric temperature and pressure graphite is the most stable form of carbon both in the bulk as well as at the nanoscale ($>5.2$ nm) \cite{Barnard2003B,Wang2005,Shenderova2002,Bundy1996,Davies1972} while diamond is stable between $\approx 1.9$ nm and $\approx 5.2$ nm \cite{Barnard2003B}. Since the utility of diamond and diamond with various color centers depends on its crystalline existence, it is imperative to study the behavior of diamond in vacuum for scientific as well as for practical purposes. Furthermore, while the determination of the size of nanoparticles using electron microscopy and dynamic light scattering are well established, their utility in levitated experiments is limited if not completely excluded. As a result it seems reasonable to devise a way by which one can determine the size of an individual levitated object while performing the experiment. This is particularly useful in experiments in which the size of a nanoparticle plays important roles. The significance of in situ size determination is further emphasized by the polydisperse nature of nanoparticles (for example see Fig. \ref{fig5}a).

In this article, we levitate high pressure high temperature (HPHT) synthesized nanodiamonds containing $\approx 500$ NV$^-$ centers (ND-NV-$100$ nm, Adamas Nanotechnology, USA) using an optical tweezer and study their behavior under different levels of vacuum. We show that as the pressure of the trapping chamber is reduced, the internal temperature ($T_i$) of a trapped nanodiamond can reach $\approx 800$ K. Due to this elevated temperature levitated nanodiamonds burn in air. We also demonstrate that the burning of nanodiamond is preventable under a nitrogen environment down to $10$ mB, but beyond that, it graphitizes. The source of heating is believed to be the absorption of $1064$ nm trapping laser light by the impurities in diamond and the amorphous carbon on the surface. Lastly, exploiting the measured damping rate of a levitated object, we present a new way of determining its size in situ.

\begin{figure}
\centering
\subfigure{
\hspace{0.5cm}
\centering
\includegraphics[width=7.50cm]{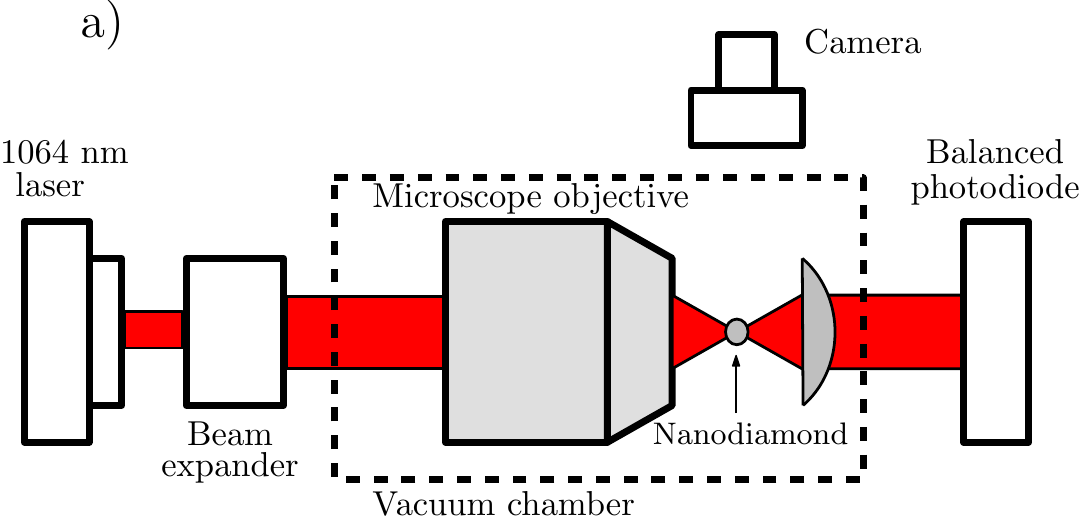}}
\subfigure{
\includegraphics[width=8.50cm]{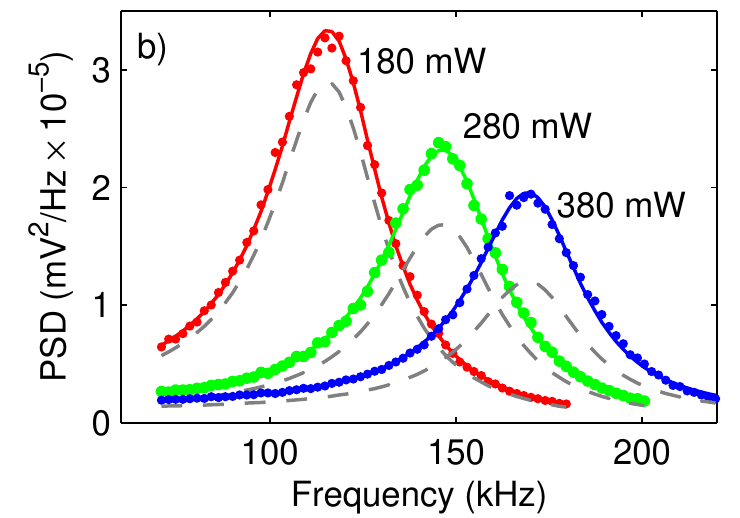}}
\subfigure{
\includegraphics[width=8.50cm]{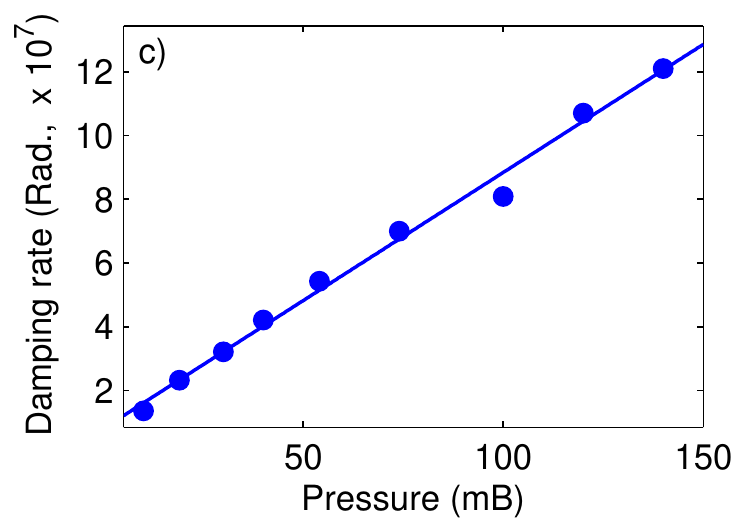}}

\caption{The trap was formed using a NA=$0.80$ microscope objective and a $1064$ nm laser. (\textbf{a}) Schematic of the experiment, and (\textbf{b}) power spectral densities (PSDs) at different trapping powers at $20$ mB along with the respective theoretical (grey dashed lines) PSDs at room temperature ($T_{CM}=300$ K). In generating theoretical PSDs, all parameters except the $T_{CM}$s have been assumed identical to the measured PSDs. Numbers besides the PSDs denote the respective trapping power at the laser focus. (\textbf{c}) Blue circles are the measured damping rate ($\gamma_{CM}$) as a function of pressure ($P$) with $180$ mW of trapping power and the blue line is the linear fit.}
\label{fig1}
\end{figure}

\section{Experimental setups}
Figure \ref{fig1}a shows a schematic of our experimental setup where we use a $0.80$ numerical aperture (NA) microscope objective to focus a $1064$ nm laser beam into a diffraction limited spot. The force resulting from the electric field gradient forms the basis of our dipole trap \cite{Gieseler2012}. The balanced photodiode visible in Fig. \ref{fig1}a provides a voltage signal generated from the interference between the directly transmitted trapping laser light and the oscillator's position dependent scattered electromagnetic radiation \cite{Gieseler2012}. Performing a Fourier transform on this voltage signal provides the measured spectral information as well as the damping rate of a levitated nanoparticle. We use this spectral information and damping rate to retrieve $T_i$ and the size of a nanodiamond. 

In the regime where the oscillation amplitude of a trapped particle is small, the trapping potential of an optical tweezer can be approximated as harmonic \cite{Gieseler2012}. Under this condition, the motion of a levitated object can be expressed as  

\begin{eqnarray}
M\frac{d^2 x}{dt^2}+M \gamma_{CM} \frac{d {x}}{d {t}}+M\omega_0^2 x&=&f(t),
\label{eqn1}
\end{eqnarray}

where $x$ is the displacement of a trapped particle from the center of the trap along the $x$-axis. $M$ and $\gamma_{CM}$, respectively, are the mass and the damping rate of a trapped particle while $\omega_0=\sqrt{\kappa/M}$ is the trap frequency and $\kappa$ is the spring constant of the trap \cite{Gieseler2012}. $f(t)$ is a Gaussian random force exerted by the gas molecules on a trapped particle with $<f(t)>\ = 0$ and $<f(t_1)f(t_2)>\hspace{0.1cm} = 2k_BT_{CM}\gamma_{CM} M \delta(t_2-t_1)$, where $k_B$ is the Boltzmann constant, $T_{CM}$ is the CM temperature of a trapped particle, and $\delta(t_2-t_1)$ is the Dirac delta function \cite{Gieseler2012}. Similar analyses for the remaining two axes are also valid. After performing a Fourier transform and rearrangement, the power spectral density (PSD) of (\ref{eqn1}) can be written as

\begin{eqnarray}
S_{x}(\omega)=\frac{2k_BT_{CM}}{M}\frac{\gamma_{CM}}{(\omega^2-\omega_0^2)^2+\gamma_{CM}^2\omega^2}.
\label{eq2}
\end{eqnarray}

We fit (\ref{eq2}) with the experimental data.

Figure \ref{fig1}b shows the PSDs corresponding to the measured voltage signals from a levitated nanodiamond for different trapping powers along with the respective fits (solid lines) of equation (\ref{eq2}) at $20$ mB. For the purpose of comparison, in Fig. \ref{fig1}b we have also included the relevant theoretical PSDs (dashed gray lines). In plotting the theoretical PSDs we have assumed that all parameters are identical to the measured PSDs except $T_{CM}$ which has been taken equal to $300$ K. Figure \ref{fig1}c demonstrates the measured damping rate as a function of pressure at a constant trapping power of $180$ mW. Later, we use this damping rate to find the size of a nanoparticle.

\begin{figure}
\centering
\subfigure{
\centering
\includegraphics[width=8.50cm]{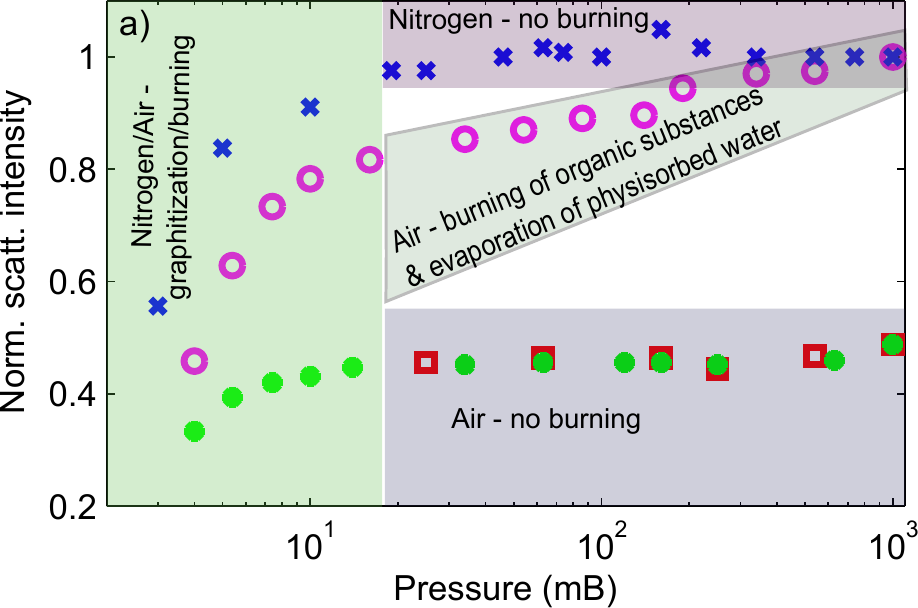}}
\subfigure{
\includegraphics[width=8.60cm]{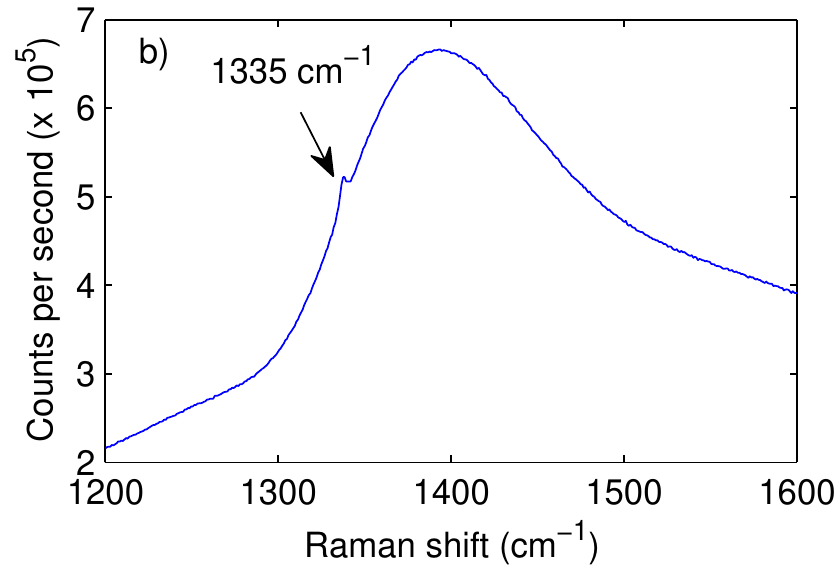}}
\caption{(\textbf{a}) Normalized scattering intensity as a function of pressure. Pink circles are for a nanodiamond as we take it to low pressures from atmospheric conditions for the first time and red squares are for the same nanodiamond but when we take it back to atmospheric pressure after keeping it at $\leq 10$ mB for about an hour. Similarly, green dots are for the same nanodiamond used in the previous two steps but when we take it to low pressures for the $2^{nd}$ time from atmospheric pressure. Trapping power was $180$ mW. Above $20$ mB physisorbed water/organic substances evaporate/burn while below this pressure diamond or amorphous carbon burns. In the second round of evacuation a nanodiamond maintains its size down to $\approx 10$ mB due to the absence of water and the organic substances on the surface. Blues crosses are the scattering intensities of a nanodiamond in a nitrogen environment. Trapping power was $\approx 300$ mW. Down to $10$ mB its size remains unchanged while below this pressure, due to elevated temperature, it graphitizes. (\textbf{b}) Raman spectrum of nanodiamonds under $785$ nm laser excitation.}
\label{fig2}
\end{figure}


\section{Levitated nanodiamonds in vacuum}
\begin{figure}
\centering
\includegraphics[width=8.50cm]{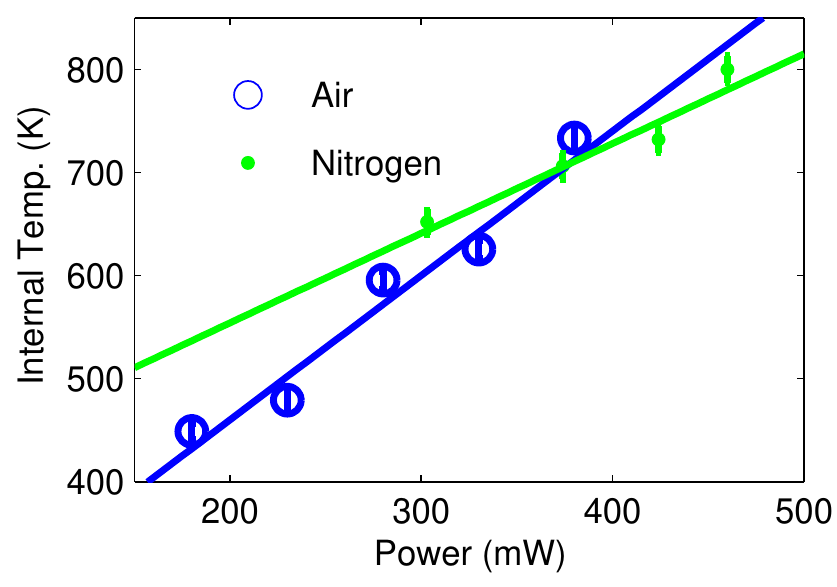}
\caption{Internal temperature ($T_i$) - blue circles in air and green dots in nitrogen at $20$ mB as a function of trapping power. Solid blue and green lines represent linear fits.}
\label{fig3}
\end{figure}

To study the behavior of diamond below atmospheric pressure, after levitating a nanodiamond with the minimum possible trapping power ($180$ mW), we gradually take it to different levels of vacuum whilst continuously monitoring its scattering intensity (size) using a camera. Figure \ref{fig2}a shows a typical plot of scattering intensity versus pressure (pink circles) from a levitated nanodiamond (for more data points see supplementary information). It can be observed that as we evacuate the trapping chamber, the scattering intensity diminishes: a levitated nanodiamond shrinks in size as the pressure is reduced. We attribute this reduction in size to the removal of physisorbed water and organic substances such as the carboxyl groups (nanodiamonds as obtained from the supplier are in water and are coated with carboxyl groups for stabilization) present on the surface of nanodiamonds down to $20$ mB where the temperature reaches $\approx 450$ K (see Fig. \ref{fig3}). Physisorbed water and organic impurities normally disappear\cite{Osswald2006} at or below $473$ K. This is further confirmed when we keep a levitated nanodiamond in a vacuum of less than $10$ mB for an extended period of time (about an hour) and take it to back to atmospheric pressure (red squares in Fig. \ref{fig2}a) and bring it down to the low pressures again. In the second round of evacuation, the scattering intensity remains constant down to $10$ mB. This unaltered scattering intensity in the second round of evacuation indicates the absence of substances which evaporate/burn at relatively lower temperatures. 

The reduction in size below $20$ mB is attributed to the burning of amorphous carbon or diamond. Amorphous carbon is generally found as an outer layer on the surface of nanodiamonds \cite{Gaebel2012,Smith2010,Osswald2006}. The burning temperature of amorphous carbon \cite{Osswald2006} at atmospheric pressure varies between $573$ K - $723$ K while the oxidation temperature of nanodiamonds\cite{Osswald2006,Gaebel2012,Xu2002} ranges from $723$ K  to $769$ K. Also, the exact oxidation temperature of nanodiamonds depend on the surface quality, the crystallographic faces, and the densities of impurities in nanodiamonds \cite{Gaebel2012,Osswald2006,Xu2002}. To confirm the presence of amorphous carbon as well as diamond in the nanoparticles that we have used in our experiments, we performed Raman spectroscopy using a $785$ nm laser. At this wavelength amorphous carbon is more sensitive than diamond \cite{Ferrari2004}. Figure \ref{fig2}b presents the relevant data. This figure clearly shows the presence of amorphous carbon and diamond peaked at $\approx 1400$ cm$^{-1}$ and at $\approx 1335$ cm$^{-1}$, respectively \cite{Smith2010,Ferrari2004,Chen1999,Merkulov1997} Given that amorphous carbon is a strongly absorbing material \cite{Nagano2008,Stagg1993,Maron1990,Duley1984}, trapping light ($1064$ nm) absorption and hence raised $T_i$ and consequent burning in an air environment is highly probable. This burning of nanodiamond in air can potentially be a major hurdle in applications where vacuum is inevitable. 

Based on the idea that an oxygen-less environment may be a cure to this problem, we have studied the behavior of levitated nanodiamonds in a nitrogen environment. This is shown in Fig. \ref{fig2}a as blue crosses for a constant trapping power of $300$ mW. It can be observed that at pressures $>10$ mB the scattering intensity hence the size of a nanodiamond remains unchanged; even though temperature is quite high (see Fig. \ref{fig3}). This is due to the fact that for burning to occur, a nanodiamond requires oxygen which is absent in a nitrogen rich environment. However, if the pressure is reduced below $10$ mB, the scattering intensity of the nanodiamond gradually diminishes. Given that there is almost no oxygen in the chamber and the reduced pressure means less cooling due to gas molecules and hence higher internal temperature, we believe this is the onset of graphitization of the nanodiamond. At atmospheric pressure graphitization of nanodiamonds starts in the temperature range $943-1073$ K and depends on the surface quality of nanodiamonds \cite{Chen1999, Xu2002}. Since we are operating at sub-atmospheric pressures, graphitization at a lower temperature is most likely to happen.

\section{Internal temperature of a levitated nanodiamond}
To verify that a levitated nanodiamond reaches such high temperatures in moderate levels of vacuum, in Fig. \ref{fig3} we present $T_i$ obtained from the same nanodiamond used in Fig. \ref{fig2} as a function of trapping power in air (blue circles) at $20$ mB. In measuring $T_i$ we have assumed that a levitated nanodiamond is at room temperature at $\approx 150$ mB. Also, since fitting uncertainties increase with the increasing pressure, $T_i$ has been plotted as a function of trapping power at a constant pressure and it was measured during the $2^{nd}$ round of evacuation at which a levitated nanodiamond maintains its size. Constancy in size/mass is a requirement of the PSD analysis. From Fig. \ref{fig3} one can see that the internal temperature reaches $\approx 750$ K at $380$ mW of trapping power in air. This is well within the reported burning temperature of amorphous carbon or diamond \cite{Xu2002,Osswald2006,Gaebel2012}. In Fig. \ref{fig3} we have also included $T_i$s obtained from a levitated nanodiamond submerged in a nitrogen environment. In this case $T_i$ reaches approximately $800$ K at the maximum trapping power. At pressures below $20$ mB, temperatures are expected to be higher given that the cooling due to gas molecules becomes less effective while the absorption remains constant. It is noteworthy that the fluorescence from NV$^-$ centers in diamond decreases significantly at temperatures beyond $550$ K and by $700$ K it reduces to $20\%$ of the room temperature value \cite{Toyli2012}. Also, at $T_i=700$ K, NV$^-$ center's fluorescence lifetime and the contrast between electron spin resonances reduce below $20\%$ of the room temperature value \cite{Toyli2012}. At a temperature above $625$ K, the spin coherence time of the NV$^-$ center decreases as well \cite{Toyli2012}. Furthermore, the highest temperature that we have measured here, using trapping powers higher than those have been used by Neukirch et al. \cite{Neukirch2015}, rules out the possibility of melting diamond as suggested in ref. $5$. Diamond usually melts at temperatures $\geq 4000$ K and requires pressure above atmospheric pressure \cite{Eggert2010}. A slight difference between the temperatures at a constant power such as at $300$ mW in Fig. \ref{fig3} between different environments can be attributed to the variation in surface qualities and the densities of impurities in different nanodiamonds \cite{Xu2002, Qian2004}. Additionally, it has been demonstrated that bigger particles heat up rapidly compared to smaller particles under the same experimental conditions \cite{MillenNat2014}. As a result, variation in the internal temperatures is expected unless all the attributes of different particles are identical. However, due to the inherent nature of levitated experiments, it is difficult to levitate particles with the same attributes in different runs of an experiment. This is further worsened by the polydispersity of nanoparticles. For example, the average size of the nanodiamonds that we have used in our experiment is quoted to be $100$ nm by the manufacturer. A representative scanning electron microscope (SEM) image of this nanodiamond is shown in Fig. \ref{fig5}a. Nanodiamonds from a few tens of nanometers to a few hundred nanometers are visible. Consequently, trapping different sizes of nanodiamonds in different runs of an experiment is possible. Nevertheless, to be consistent throughout the experiment, we levitate nanodiamonds of similar size by monitoring their scattering intensities. Also, next we present a way of determining the size of an individual levitated object from the measured damping rate ($\gamma_{CM}$) that it encounters while oscillating inside the trap. For the purpose of following calculations, we assume that a levitated nanodiamond is of spherical shape.

\section{Determination of the size of a levitated nanodiamond}
The effective damping rate as shown in Fig. \ref{fig1}c can be expressed as $\gamma_{CM}=\gamma_{imp}+\gamma_{em}$, where $\gamma_{imp}$ and $\gamma_{em}$ are the damping rates due to the impinging and emerging gas molecules, respectively \cite{MillenNat2014}. $\gamma_{imp}$ can be written as $\frac{4\pi}{3}\frac{m N R^2\bar{v}_{{imp}}}{M}$ while $\gamma_{em}$ is related to $\gamma_{imp}$ by $\gamma_{em}=\frac{\pi}{8}\sqrt{\frac{T_{em}}{T_{imp}}}\gamma_{imp}$, where $R$, $N$, $m$, and $\bar{v}_{{imp}}=\sqrt{\frac{8k_BT_{imp}}{\pi m}}$ are the radius of a trapped particle, the number density of gas molecules at pressure $P$, molecular mass, and the mean thermal velocity of impinging gas molecules, respectively \cite{MillenNat2014}. $N$ can further be expressed as $N=N_0P/P_0$, where $N_0$ is the number of gas molecules per cubic meter at atmospheric conditions and $P_0$ is the atmospheric pressure. On substitutions of various terms, one can express $R$ as

\begin{figure}
\subfigure{
\centering
\hspace{1.25cm}\includegraphics[width=6.7500cm]{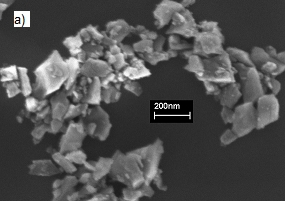}}
\subfigure{
\centering
\includegraphics[width=8.500cm]{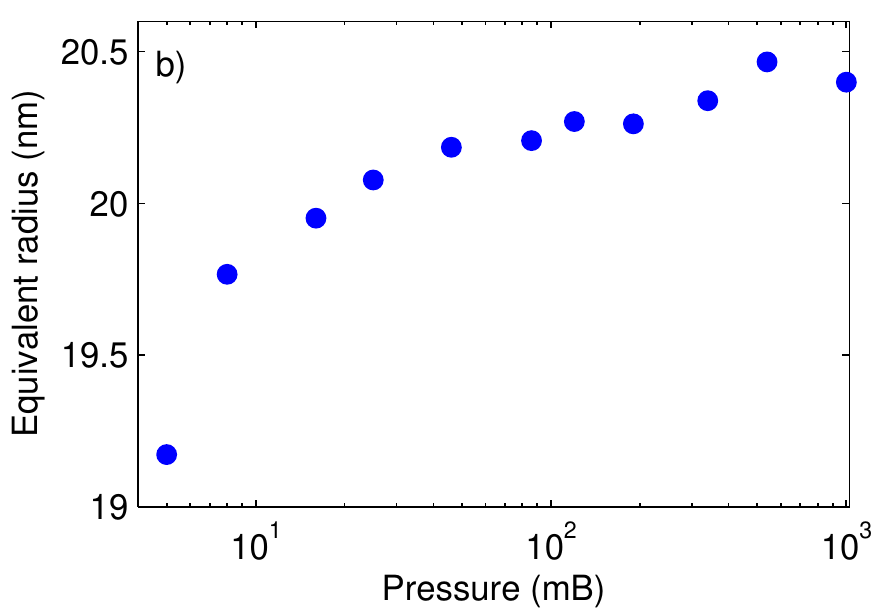}}
\caption{(\textbf{a}) Scanning electron microscope image of nanodiamonds as received from Adamas Nanotechnologis Inc., USA, and (\textbf{b}) the equivalent radius using equations (\ref{eq18}) and (\ref{eq22}) of the trapped nanodiamond for which the internal temperatures were found in Fig. \ref{fig3} in air as a function of pressure.}
\label{fig5}
\end{figure}

\begin{eqnarray}
R&=&\frac{m\bar{v}_{{imp}}N}{\rho\gamma_{cm}}(1+\frac{\pi}{8}\sqrt{\frac{T_{em}}{T_{imp}}}),
\label{eq18}
\end{eqnarray}
where $M$ has been expressed as $\frac{4}{3}\pi R^3\rho$ and $\rho$ is the mass density of diamond.     

Given that the levitated nanodiamonds burn, equation (\ref{eq18}) gives the ultimate size of a nanodiamond for which we previously found temperatures. That is, it is the size of the nanodiamond after the first round of evacuation. The actual size of a nanodiamond before burning can be found using scattering theory. The scattering intensity of a Rayleigh particle ($R<<\lambda$) is given by $I_{s}=\frac{8\pi}{3}k^4R^6|\frac{\epsilon -1}{\epsilon +2}|^2 I$, where $k=\frac{2\pi}{\lambda}$ and $I$ is the intensity of the trapping light \cite{BohrenHuffman}. Provided that we know the scattering intensity (see Fig. \ref{fig2}b) at different pressures, we can find the actual size of a nanodiamond using equation (\ref{eq22}): 

\begin{eqnarray}
R_p=(\frac{I_{s_p}}{I_{s}})^{1/6}R,
\label{eq22}
\end{eqnarray}
where $R_p$ and $I_{s_p}$ are the radius and the scattering intensity of the particle at pressure $P$, respectively.

As examples, using the model developed here, we estimate the sizes of the nanodiamonds for which we have presented internal temperatures in Fig. \ref{fig3}. Using equations (\ref{eq18}) and (\ref{eq22}), and parameters $N_0=2.43\times 10^{25}$, $T_{imp}=300$ K, $T_{em}=450$ K, $\rho=3500$ kg/$m^3$, $m=4.81\times 10^{-26}$ kg , $P=20$ mB and $\gamma_{cm}=2.18\times 10^5$ radian with the minimum trapping power of $180$ mW, Fig. \ref{fig5}b shows the radius of the trapped nanodiamond at various pressures in air. It can be observed that when the nanodiamond was initially trapped at atmospheric pressure, its diameter was $\approx 41$ nm. Similarly, for the nitrogen case using the same parameters except $\gamma_{cm}=2.22\times 10^5$ radian and $T_{i}=650$ K, we get the ultimate diameter of the nanodiamond is $\approx 38$ nm. Given the uncertainty in the shape of nanodiamonds as visible in Fig. \ref{fig5}a, the nanodiamonds that we have used to find $T_i$s in air and nitrogen ambients are of similar size. This is also in good agreement with the technique (initial scattering intensities) that we have utilized to trap similar size nanodiamonds in different runs of an experiment. Furthermore, even though the actual dimensions of a nanodiamond will be different from $R$ due to its asymmetric shape, the estimated size provided by our model is well within the distribution visible in the SEM image (Fig. \ref{fig5}a). Lastly, we believe that the method developed here for the determination of size of an individual particle can be used in any levitated experiment.

\section{Discussion}
We have demonstrated that nanodiamonds burn in air while they graphitize in a nitrogen ambient by absorbing trapping laser ($1064$ nm) light as the cooling due to gas molecules becomes less effective with decreasing pressure. We believe that amorphous carbon, a strongly absorbing material, present on the surface of nanodiamonds is a key reason for this. We also think that purer nanodiamonds instead of the currently available HPHT synthesized nanodiamonds can be a cure to this problem. Our Brownian motion based analysis has shown that the internal temperature of a levitated nanodiamond can reach up to $800$ K. This rules out the possibility of melting diamond which requires \cite{Eggert2010} a temperature $\geq 4000$ K. Lastly, exploiting the damping rate that a particle encounters while in motion, we have developed a new way of determining its size. We consider that this new technique will be useful in present and future levitated experiments where the traditional electron microscopy and dynamic light scattering based size determinations are not suitable.

\section{Method}
Nanodiamonds containing $\approx 500$ NV$^-$ centres (ND-NV-$100$ nm) were bought from Adamas Nanotechnology Inc, USA. The average size of the nanodiamonds quoted by the manufacturer is $100$ nm. To prevent agglomeration we sonicate as received nanodiamonds for $\approx 10$ minutes in an ultrsonic bath and then put them into a nebulizer and inject them into the trapping chamber. The trapping chamber is continuously monitored by a CMOS camera (Thorlabs Inc). Once a nanodiamond is trapped, the trapping chamber is evacuated to study the behaviour of nanodiamonds in vacuum. Power spectral density data were collected using a  balanced photodiode (Thorlabs Inc) and a Picoscope oscilloscope (Pico Technology, UK). In the case of nanodiamonds immersed in nitrogen, the trapping chamber was purged with nitrogen fifteen times.

\bibliographystyle{naturemag}

\end{document}